\documentclass[preprint]{aastex} 
\newcommand{{\HII}}{H\,{\sc ii}}

\newcommand{\um}{\,$\mu$m}
\newcommand{\kms}{\,km\,s$^{-1}$}
\newcommand{\wno}{\,cm$^{-1}$}
\newcommand{\bra}{Br\,$\alpha$}

\shorttitle{[S\,IV] in NGC 5253}
\shortauthors{Beck et al}

\begin{document}

\title{[S\,IV] in the NGC 5253 Supernebula: Ionized Gas Kinematics at  High Resolution  \\
   }

\author{Sara C. Beck \altaffilmark{1,2}, John H. Lacy\altaffilmark{2,3}, Jean L. Turner\altaffilmark{4},  Andrew Kruger\altaffilmark{5,2}, Matt Richter\altaffilmark{5,2}, Lucian P. Crosthwaite\altaffilmark{6} }
\altaffiltext{1} {Department of Physics and Astronomy, Tel Aviv University, Ramat Aviv, Israel email: sara@wise.tau.ac.il}
\altaffiltext{2}{Visiting Astronomer at the Infrared Telescope Facility, which is operated by the University of Hawaii under Cooperative Agreement no. NNX-08AE38A with the National Aeronautics and Space Administration, Science Mission Directorate, Planetary Astronomy Program.}
\altaffiltext{3}{Department of Astronomy, University of Texas at Austin, Austin Tx 78712}
\altaffiltext{4}{Department of Physics and Astronomy, UCLA, Los Angeles, CA 90095-1547}
\altaffiltext{5}{Department of Physics, University of California at Davis, Davis CA 95616}
\altaffiltext{6} {Northrop Grumman Aerospace Systems, San Diego, CA 92127}

\begin{abstract}
  The nearby dwarf starburst galaxy NGC 5253 hosts a deeply embedded radio-infrared supernebula excited by thousands of O stars. We have observed this source in the $10.5\mu$m line of $S^{+3}$ at 3.8\kms~spectral and {1.4\arcsec}  spatial resolution, using the  high resolution spectrometer TEXES on the IRTF.   The line profile  cannot be fit well by a single Gaussian.    The best simple fit describes the gas  with two Gaussians, one near the galactic velocity with FWHM 33.6\kms~and another of similiar strength and FWHM 94\kms~centered $\sim20$\kms~to the blue.  This suggests a model for the supernebula in which gas flows towards us out of the molecular cloud, as in a ''blister'' or ''champagne flow'' or  in the   {\HII} regions  modelled by \citet{ZH}.    
   \end{abstract}


\section{Introduction}
Super star clusters are a common mode of star formation in starburst galaxies.  These star clusters form deeply embedded in molecular clouds, and the young stars disperse the remaining cloud material to become visible clusters, while contributing to the ionization, mechanical energy, and metal enrichment of the starburst.  Through the course of their short lifetimes, O and B stars have the potential to massively change their environments through the action of stellar winds and supernovae, which could have the effect of either halting further star formation, or inducing a new round of cluster formation. What are the observed effects of a young super star cluster on its immediate environment? To address this question, we are carrying out observations of the gas kinematics of the supernebulae these star clusters excite. This paper examines the gas kinematics in one of the closest such supernebulae, which is also the most luminous known. 

The dwarf galaxy NGC 5253 has created towards its center dozens of young, bright super star clusters (SSCs) with ages of a few to tens of Myr  \citep{MH95,1997AJ....114.1834C, 2004ApJ...603..503H}.  In the same area as the young clusters is an infalling streamer of CO \citep{TBH97,MT02}) and H\,I \citep{2008AJ....135..527K,LS07} and coincident with it, a  filament of emission first detected in [O\,III] by \citet{1981PASP...93..552G} and lately interpreted by \citet{2011ApJ...741L..17Z} as an ``ionization cone".  It is possible that the infalling molecular stream encouraged formation of the clusters, and that the cone is the outside of the same cold gas streamer, lit up by the ionization.  This illustrates how  clusters can influence the galaxy well beyond their immediate vicinity.  

Infrared observations \citep{TB03,AAH04} of the core of the starburst region find a 
deeply embedded star cluster located $0.3-0.4\arcsec$ (5-6 pc at 3.8 Mpc) to the northwest of the brightest optical cluster (cluster ``5" of
\citet{1997AJ....114.1834C}.) The embedded cluster is not detected at wavelengths shorter than 1.9\um\ due to  extinction, which is estimated to be  $\rm A_V\sim 17$, and appears to originate within the source \citep{TB03,
2005A&A...429..449M}.  The embedded cluster is coincident with bright Brackett line emission associated with
a strong radio source, the ``supernebula" \citep{1998AJ....116.1212T, 2000ApJ...532L.109T,TB04}.  The supernebula requires the ionization equivalent of 1200 O7 stars to excite the dense, parsec-sized core of the H\,II region \citep{TB04,2001ApJ...557..659M}.  In addition to the dense core, there is a more extended halo to the \HII\ region, requiring the excitation of 2000 O7 stars within the central 5 pc, and up to 7000 O stars within a radius of 20 pc \citep{TB04}.  The
extinction, the high inferred density of several thousand per cm$^{3}$ for the \HII\ region \citep{TB04, MI10} and the young spectroscopic age for the cluster \citep{AAH04}
suggest that this is one of the youngest known superclusters, still in its embedded phase and 
possibly still in the process of formation.

The current star forming episode in NGC 5253 is dominated by the ``supernebula",
which is only $\sim$1 pc in radius. A luminosity in
young stars of $\sim 10^9~\rm L_\odot$ and ionization rate of $\sim 4$--$7 \times 10^{52}~\rm s^{-1}$ within a 2--10 pc region require the presence of a stellar cluster containing thousands of O stars and a dense, 
dusty, compact \HII\ region. This is not a static configuration: hot ionized gas will expand under its high pressure, and young stars, especially O stars,  are active and drive powerful winds and outflows.   What effect is this young, forming super star cluster having on  its
surroundings?  Are stellar winds evident? If there are O stars present, especially evolved O stars \citep{AAH04}, then why is this
 H\,II region still embedded? What can the gas kinematics be in such an extreme {\HII} region?

The kinematics of the supernebula have been observed in the near-infrared Brackett  recombination lines of H\,I \citep{TB03} . 
The FWHM of the near-infrared lines, measured with 12\kms\ resolution, was $76\pm2.5$\kms\ over the central few arcseconds surrounding the 
supernebula. These velocities are not much higher than the FWHM of some individual Galactic  compact {\HII} regions excited by one or a few O stars \citep{B08}. But compared to a normal  \HII\ region, the supernebula has thousands of times more
stellar mass in roughly the same volume.  This led \citet{TB03} to suggest that the expansion velocities may actually be less than the escape velocity for much of the cluster, so the supernebula could be gravitationally bound; a giant photosphere, rather than a conventional {\HII} region.  This would correspond to the ``Type III'' cluster of
\citet{2002MNRAS.336.1188K}, in which the escape velocity exceeds the sound speed. 
On somewhat larger (60-100 pc) scales, gas motions have also been observed through the H53$\alpha$ radio recombination line  \citep{RG07}, and in red lines of H\,$\alpha$, [N\,II] and [S\,II] 
\citep{MI10}. 
The H53$\alpha$ line was measured with 44\kms\ resolution and had an 
observed FWHM $\sim100\pm12$\kms\ over 
the inner 4\arcsec\ ($\sim$60 pc, see Figure 5 of \citet{RG07}).  %%
 \citet{RG07} found a gradient of $10~\rm km~s^{-1}~arcsec^{-1}$ over this region, which they interpreted as rotation. \citet{MI10} also find a gradient, over scales similar
 to the H53$\alpha$ image, but it is a gradient in differential velocity between
 the forbidden lines [N\,II] and [S\,II] and H\,$\alpha$. Neither the H53$\alpha$ nor the red lines are ideal for probing spatial and kinematic effects in the
 the supernebula; for the RRL the beam is too big (2\arcsec ) to distinguish
 motions within the nebula while extinction within the supernebula prevents observation in the 
 optical.

The strong mid-infrared emission lines of the metal ions  are useful probes of ionized gas in obscured regions.  They are relatively free of extinction compared to visual lines and the low thermal broadening due to the ionic weight means that they are much better at probing kinematics than is hydrogen \citep{ZL08,DJ03}.   At typical \HII~region electron temperature of $10^{4}$ K the thermal width of a hydrogen line will be 21\kms\ FWHM, but the thermal width $V_m$ of line emitted by a metal of mass $m_m$ will be smaller than that of a hydrogen line by the ratio $V_m/V_h  = (m_h/m_m)^{1/2}$.  
For NGC 5253, which is a high-excitation source,  the strongest line at mid-infrared wavelengths that can be observed from the ground \citep{BL96} is [S\,IV] 10.5\um.  This promises a factor of  5.66 reduction in the thermal component of the line profile. 
We accordingly observed [S\,IV] 10.5\um\ from the supernebula in NGC 5253 with high spectral and $\sim1\arcsec$ spatial resolution, with the aim of separating the thermal component of the line width and determining the true bulk motions of the gas.  

\section{Observations; the TEXES [S\,IV] Data Cube}

NGC 5253 was observed on the night of 4 June 2010 with  TEXES, the Texas Echelon Cross Echelle Spectrograph \citep{LA02}  spectrometer,  on the NASA IRTF on Mauna Kea.     TEXES is a sensitive spectrometer for the 5--25\um\ region, with three resolution modes: these data were obtained in the high resolution mode,  which gives spectral resolving power $R\sim80,000$.   The slit was 29 pixels long and the plate scale $0.36\arcsec\times0.95~\rm 
km~s^{-1}$ per pixel. Diffraction and seeing limit the final spatial resolution to 1.4\arcsec\ and the spectral resolution to $\sim3.8$\kms.  We offset from the guide star to the position of the radio supernebula \citep{TB04}. The slit, which was oriented NS, was stepped across the galaxy  in {1\arcsec} increments, first west, then back to the center to check the pointing, which was excellent, and then east.   At each position the galaxy was nodded 12\arcsec\  north, taking the emission region off the 9\arcsec\ slit.   The beams were subtracted and the spectra were combined to create a data cube.

Wavelength calibration is done by reference to Earth atmospheric lines seen in the spectrum of the source and checked against the spectrum of the asteroid Vesta, which was also used as the divisor to remove the atmospheric transmission.  There are no strong lines close to the wavelength of [S\,IV] in NGC 5253, which from the redshift of $\sim400$ \kms\ and the [S\,IV] rest wavelength of 951.43\wno\ is predicted to be 950.16\wno.  Velocities quoted are heliocentric. 

The total flux in the [S\,IV] map (the zeroth moment of the data cube) is $6\times10^{-12}~\rm
 erg~s^{-1}~cm^{-2}$, 50\% higher than the previous result of \citet{BL96}. This may reflect a more thorough mapping of the [S\,IV] emitting region or an offset in the
calibration.   This flux agrees, within the calibration uncertainties, with the total [S\,IV] emission seen by the Spitzer IRS in a much larger slit.  

 \section{Spatial Distribution of the Gas \label{kinematics}}
 The TEXES data give us effectively a spectrum at each point in the sky.  The data were obtained with the slit oriented NS, so by collapsing the spectra along the slit we obtain 1-d spectra showing the line profile in each EW position.  The line was detected in each position over the total spatial extent of $5\arcsec$, but the strongest position is more than 5 times stronger than any other. The spectra are displayed in Figure 1.  The strongest position is  $1\arcsec$ east of the point arrived at by offset from the guide star to the radio source. In light of the spatial resolution, beam size, and usual offset accuracy this is not significant and we identify the peak of the [S\,IV] emission with the radio supernebula at : R.A.$= 13^h39^m55^s.96$, decl. $= -31^{\circ}38'24.38\arcsec\ $  \citep{TB04}.  In Figure 2 we show the [S\,IV] data cube in the form of a grid of spectra, obtained by binning the data into $1\arcsec~\times~1\arcsec$ cells.  The grid steps are comparable to the diffraction limited beam size of the [S\,IV] data.  
 
 The [S\,IV] is concentrated but it is not point-like. In Figure 3 we show  the spatial distribution of the [S\,IV] emission integrated over the whole velocity range (the zeroth moment of the data cube). The [S\,IV] emission is  symmetric and only slightly extended EW, but in declination it is clearly extended and asymmetrically so: the emission falls off sharply  north of the peak but extends  $3\arcsec$ towards the south.  This is seen along individual NS slits  and is not a result of the observing procedure or the data reduction.  The extension of the [S\,IV] line emission south of the supernebula agrees with the high spatial resolution radio continuum map of  \citet{TB04}. Their image shows a ``filamentary arc" of free-free radio emission extending south
 of the main source and appearing to curve around the second nuclear  star cluster (the less obscured cluster which is optically visible but not a strong radio source); they suggest that it may be a blister feature or cometary \HII\ region.
 
\section{The [S\,IV] Line Profile}
It is clear from Figure 1 that the [S\,IV] line profile is not simple and symmetric. There is obvious red-blue asymmetry around the peak, with the blue side extending to higher velocity than the red, and the whole line profile appears distinctly non-Gaussian.  Given the complex line shape, and that there are no models for the gas kinematics of a supernebula or for what the resulting line profile `should' look like, caution is necessary in trying to fit the line profile.  While it is possible to obtain a fit of great formal goodness by using many components, it may not be physically meaningful.  Here we concentrate on the simplest fits. They are not formally perfect, but can still reveal a great deal about the supernebula. 

 Since we do not a priori know the kinematics of gas in the supernebula, the first step is to test what aspects can and cannot be described by the simplest possible fit, which is a single Gaussian.  We used the  IDL procedure LMFIT to search for the fits.  The best single Gaussian fit to the [S\,IV] line, and the formal uncertainties to the fit, are:  center at $+377\pm0.5$\kms\ and $\sigma=27.8\pm0.6$\kms , or FWHM of 65\kms.  This fit  and the residuals are shown in Figure 4.  The residuals in Figure 5 clearly show that two features of the [S\,IV] profile are not well fit by a single Gaussian: a narrow component about 10\kms\ red of the main peak, and excess emission on the blue side extending out to -100\kms.  
 
 Two Gaussians, shown in Figure 5, produce a much better formal fit.  One Gaussian is centered very close to the galactic velocity at  $391\pm0.8$\kms\ and has
$\sigma=14.1\pm0.7$\kms\ or FHWM 33.6\kms.   The second, of almost equal peak strength,  is  offset $~17\pm6$\kms\ to the blue and has $\sigma=39.25\pm1.7$\kms\ or FWHM 94\kms.  The formal reduced $\chi^2$  of the single Gaussian fit is 2.04 and of the two Gaussian fit, 1.29. The absolute values of $\chi^2$ depend on the noise level and may have systematic errors; the relative values of $\chi^2$ for the  fits are more significant, and agree with the clearly superior appearance of the two Gaussian fit and its residuals.  We use these fits as guides for our discussion of the nebular structure; as more information on the structure and kinematics of the supernebula, the cluster and its environment is recorded, a more realistic model can emerge.   
 
Gaussian line shapes are produced by random motions.   The gas in the supernebula is certainly undergoing at least two kinds of random motions: thermal, which depend on T$_e$,  and virial turbulence  which depends on the gravitational potential of the star cluster mass.  There may be other sources of random motions and line broadening present. For example, the intense star formation that created the embedded cluster was presumably accompanied by stellar winds and outflows that could have generated significant turbulence.  But these other sources, if they exist, are not yet quantifiable. We therefore discuss here only the gravitational and thermal effects.  

  \subsection{Gravitational Effects }
 How do the velocities from the fit compare to estimates of the gravitational energy of the system?  From the N$_{Lyc}$ and total luminosity of the supernebula and a Kroupa IMF extending to 0.1M$_\odot$, the embedded cluster has total mass $\sim3\times10^5$M$_\odot$.  Using the virial velocity dispersion $\sigma_r=({GM_{cl}/ 3R})^{1/2}$
and assuming that the 
Gaussian  component of the line emission originates in the same compact region as the free-free continuum,  which we take from the image of \citet{TB04} to be R $\sim1$pc, gives $\sigma_r\sim$20\kms\ or FWHM for the bulk virial motions $\sim47$\kms.   This rough calculation shows that the bulk of the gas has velocities consistent with other estimates of the total cluster mass; there is no need for a truncated or top-heavy IMF to explain the M/L ratio.

  \subsection{Thermal Broadening and the Profiles of [S\,IV] and Brackett Lines} 
Any nebular line will have a thermal width which depends on the ionic weight and T$_e$. A motivation to observe the  [S\,IV] line is that sulfur's weight makes it less susceptible to thermal broadening  than are hydrogen lines, so it is possible to see the gas motions in more detail.  How does the [S\,IV] line compare to  the Brackett line profiles of \citet{TB03}? We will concentrate on the \bra\ line  because of the longer wavelength and lesser susceptibility to extinction, since A$_{2\mu m}\sim2$ mag, and we sum the
\bra\ over 1\arcsec\ to match the [S\,IV] slit. The best single Gaussian fit to the
\bra\ line is centered at 377\kms\ and has  
$\sigma=31.5$\kms , FWHM = 74\kms. 
The best two Gaussian fit has a narrow component, 
$\sigma=22.7$\kms ,  centered at 391\kms\  and a blue component with $\sigma=49.1$\kms\  centered at 368\kms.  The best single Gaussian fit  and residuals are shown in Figure 6 and the best two Gaussian fit and residuals in Figure 7.  For \bra\ the single Gaussian and two Gaussian  fits have the same reduced $\chi^2\sim0.96$.  This is a striking contrast to the [S\,IV] line, which could clearly not be fit well by one Gaussian.  

The [S\,IV] and \bra\ lines are expected to probe the same body of gas; while  S$^{+3}$ and H$^+$ have different ionization potentials they are expected to co-exist in a nebula ionized by hot stars because almost all the sulfur will be in S$^{+3}$.  This is supported by the ionic abundance of S$^{+3}$ derived from the [S\,IV] line strength, which is at least as high as the elemental abundance of sulfur in NGC 5253 (assumed to be $\sim$0.29 solar, \citet{1997ApJ...477..679K}'s value  for the galactic metallicity).   So the lines should resemble each other except in so far as the Brackett lines have lower spectral resolution, greater thermal broadening and more sensitivity to extinction.  That two Gaussian fits match both the [S\,IV] and \bra\ profiles show a basic agreement of the two lines. There is more evidence for the blue wing in [S\,IV] than in \bra , even degrading the [S\,IV] resolution to 32\kms\ (the sum in quadrature of the instrumental resolution and the thermal width of H at $2\times10^4$ K, see next section), leaves more blue emission than is seen in the \bra.  But the \bra\ line is affected by a negative feature  70\kms\ blue of the peak, perhaps imperfectly cancelled atmospheric absorption, which could disguise a blue excess.  We would need better \bra\ data to be sure that the line profiles are consistent.

\subsubsection{T$_e$ in the Supernebula}
 One can derive the thermal broadening of a spectral line, and thus T$_e$ of a nebula, by comparing the line widths of ions of different weights.  We use for this the single Gaussian fits; although the single Gaussian is not an optimum fit for the [S\,IV] it  is most appropriate for the current \bra\ line data. If higher resolution Brackett spectra become available, it may be possible to fit the line better and derive T$_e$ more accurately. 
 
 If we compare the single Gaussian line fits for hydrogen and sulfur and assume that the widths are due to the convolution of a thermal width, which is 5.66 times greater for hydrogen than for sulfur, with a turbulent or other random width which is the same for both species, the resulting simultaneous equations can be solved to give a thermal width of 5.8\kms\ for sulfur and 32.5\kms\ for hydrogen.  This is significantly higher than the 21\kms\ FWHM velocity broadening associated with a `normal' \HII\ region temperature  of $10^{4}$\,K and requires  T$_e$ of $2.5\times10^{4}$\,K. 
   This T$_e$ value is quite realistic for the supernebula;  T$_e$ depends on ionization parameter $U$, and the small size and intense radiation of the supernebula create a  very high ionization parameter,  $log~U\sim-0.1$.   \citet{DS91} calculate cooling in ionized clouds and predict a {\it minimum} nebular temperature of $2\times10^4$\,K for this $U$ and the metallicity of NGC 5253.

 \subsection{Modelling the Blue Wing} 
Both the single Gaussian and two Gaussian fits indicate  excess blue emission.  Blue wings are often observed in the spectra of embedded, highly obscured stars and clusters.  An obvious model for this profile in a {\HII} region is that the blue wings are  due to supression of the red side by extinction from absorbing  dust mixed with the emitting gas in a symmetric expanding flow. ${\rm A_{[S\,IV]}\sim 0.7A_K}$, which would give A$_{[S\,IV]}$ of 1.4 magnitudes: a significant  amount of obscuration, worse than the \bra\ suffers. But it should be noted that the silicate feature is weak in NGC 5253 and this figure  is an upper limit, possibly quite a high one, for the [S\,IV] obscuration.  
But while there is ample extinction in the supernebula   \citep[][found A$_{4.05\mu m}$ to be 0.8 mag]{TB03} the model is not very physically plausible.  First is a lifetime problem:  if the supernebula were really expanding at $\sim100$\kms ,  as the linewidths would imply in this model,  it would have achieved the radio size of 0.7\,pc in
$\sim6\times10^3$ years and  the 1\arcsec\ of the [S\,IV] resolution in only $1\times10^5$ years old. Yet observations of the  star cluster and the evidence for possible WR activity in the area suggest that it is $\sim 3$\,Myr old \citep{AAH04}. The second argument against this model is the line shape itself.  If the absorbing material in the expanding cloud is uniformly distributed, then extinction sufficient to reduce severely the red-shifted emission would be expected to distort and shift the line center relative to other lines with different obscuration.  But the peak velocity of the [S\,IV] agrees with the Brackett lines \citep{TB03}, which agree with each other, in spite of their different extinctions.

Another possible explanation of the blue wing is that  the emitting region contains a secondary source or sub-cluster blue-shifted relative to the main cluster.   If so,  the line shape should vary with position, except in the unlikely case that the secondary source is exactly in the same line of sight as the main cluster.   We analyzed the spatial dependance of the line shapes by interpolating onto a grid $0.36\times0.33 \arcsec$ pixels (close to the $0.36\arcsec\ $ size of a detector pixel). In Figure 8 we show the spectra obtained in the peak pixel and in a $3\times3$ pixel box (minus the peak pixel), a $5\times5$ minus the inner $3\times3$ pixels,  $7\times7$ minus the inner $5\times5$ and $9\times9$ minus the inner $7\times7$ pixel boxes surrounding it.  To the limits of the signal-to-noise, no variation is apparent. We believe that while a secondary blue-shifted source cannot be ruled out, there is at present no evidence for it.   It should be noted that high resolution radio maps \citep{2000ApJ...532L.109T} do find another small source $0.3\arcsec$ from the main cluster, but it is a much smaller fraction of the total radio emission than the blue wing is of the [S\,IV] flux.  

\subsubsection{Instrinsically Asymmetric Motions--A Blue Flow?}
The supernebula line profile is not unique: \citet{2007AJ....133..757H} fit Brackett lines in the super star cluster nebulae in
He 2-10 with narrow ($60-71$\kms ) plus broad ($200-300$\kms )  profiles; the broad
profiles are blueshifted with respect to the narrow components, which they attribute
to flows in an aging starburst. 
\citet{GG07} report Br$\gamma$ lines with FWHM $60-110$\kms~and non-Gaussian wings up to $\sim200$\kms\ wide for 17 embedded cluster sources in the Antennae galaxies, also in regions that are $\sim 100$\,pc extent, several times larger than those we consider here. Since supernebulae are in many respects (e.g. radio spectrum,  evolutionary stage)  scaled-up version of Galactic Ultra-Compact { \HII} regions, and  \citet{B08} showed that the H\,I recombination line profiles of embedded clusters resemble the profiles obtained by superposing the H\,I emission of many ultra-compact{\HII} regions,   it is natural to look at these { \HII} regions--which can be studied in great spatial detail--for clues to the possible gas motions. 

The infrared emission lines of UC{\HII} regions frequently have a narrow line center and wide low level flux, over  velocity ranges like those of  the [S\,IV] lines, and are asymmetric, in the same sense  (blue excess) as we see in NGC 5253.   
\citet{PG84}  present \bra\ spectra of embedded young stellar objects which have deconvolved FWHM $50-235$\kms\ and, in some cases, broad blue wings. \citet{SK05} shows radio recombination lines of similar FWHM from hyper-compact and ultra-compact~ {\HII} regions. High resolution spectral mapping \citep{ZH} and modeling has shown that in these cases the ionizing star(s) lie on the near side of the natal molecular cloud and that gas flowing along the cloud surface and away from the cloud creates the blue wing. 

We think this is the most likely picture for the supernebula and the most natural explanation of the line shapes. It explains why the core of the line is almost symmetric; that gas has not entered the flow zone and is behaving like a normal {\HII} region.   We note that although Wolf-Rayet winds are not thought to turn on until $\sim3$~Myr, \citet{ZL08} found that most observed ultra-compact~{\HII} regions show evidence of the influence of stellar winds.

\section{Discussion and Conclusions}
The supernebula in NGC 5253 is the most compact embedded super star cluster known, one of the most obscured, and one of the very brightest. It is also probably the youngest, as shown by its small size, high gas density \citep[about ${\rm 3-4\times10^4 cm^{-3}}$,][]{TB04}, and the very high ionization state of the gas (\citet{CB99} arrived at an age of 2-3 Myr from analysis of the radiation field, agreeing with 3 Myr  \citet{AAH04} found from the near-infrared continuum).  It is in a complex region of many star clusters and molecular gas. As it ages,  emerges from its embedding material, and produces supernovae it will affect its surroundings, and indeed the whole galaxy, profoundly.  The kinematic information gained from the [S\,IV] line has let us probe the gas motions in more detail than ever and can point out how the supernebula may evolve. 

The [S\,IV] spectrum we present here measures  the ionized gas excited by an embedded star cluster with the highest spectral resolution yet achieved.  The [S\,IV] line shapes are quite constant throughout the [S\,IV] emission region, with no sign of the apparent gradient seen on larger scales by the radio recombination lines \citep{RG07}, 
 nor those seen in the red forbidden lines of [S\,II] and [N\,II] \citep{MI10}.   
The line profile is not a simple Gaussian, but is asymmetric about the peak with a blue wing.   We look at the simplest possible fits as starting points for discussion of the kinematics. The single Gaussian which is closest to fitting the line has FWHM 65\kms; this fit leaves excess blue emission and an additional narrow (22\kms ) redshifted peak.  
A much better formal fit is obtained with two Gaussians, one of FWHM 33.6\kms~at galactic velocity and another with FWHM 94\kms~offset $17\pm6$\kms~to the blue.  There are no clear features in the residuals of this fit. 

The best formal fit leads to a model in which the gas may be viewed as either still confined to the cluster (the main peak), or having entered the flow away from the cluster (the blue component).  In reality these are not two distinct bodies of gas, but continuous.  In this model the supernebula's kinematics are like those of a blister {\HII} region, where the ionized gas flows out of the cloud.  Note that when \citet{TB04} spatially resolved the ionized gas of the supernebulae, they found an arc or cometary shape,  which again resembles a  blister {\HII} region.    As more information on the supernebula and on the surrounding clouds becomes known,  it should be possible to refine the fits according to what is physically most realistic, as well as formally good.    

This blister or pressure-driven flow picture is an attractive one for supernebula.  Because the ionized gas does not simply expand, there is no life-time problem.  The overpressure of the ionized gas is relieved by the outflow, while the supernebula maintains its small size, and this can continue as long as the gas flowing out is replenished by ionization of molecular gas from the cloud.   Other extragalactic supernebula should be observed via the infrared metal lines to determine if they show similar line profiles at high spectral resolution. 
\acknowledgements
  TEXES observations at the IRTF were supported by NSF AST-0607312 and by AST-0708074 to Matt Richter. This research has made use of the NASA\&IPAC Extragalactic Database (NED) which is operated by the Jet Propulsion Laboratory, Caltech, under contract with NASA.  SCB thanks Mike Shull and the University of Colorado for hospitality and useful discussions while part of this work was done.

\clearpage

\begin{figure}
\begin{center}
\epsscale{1.4}
\plottwo{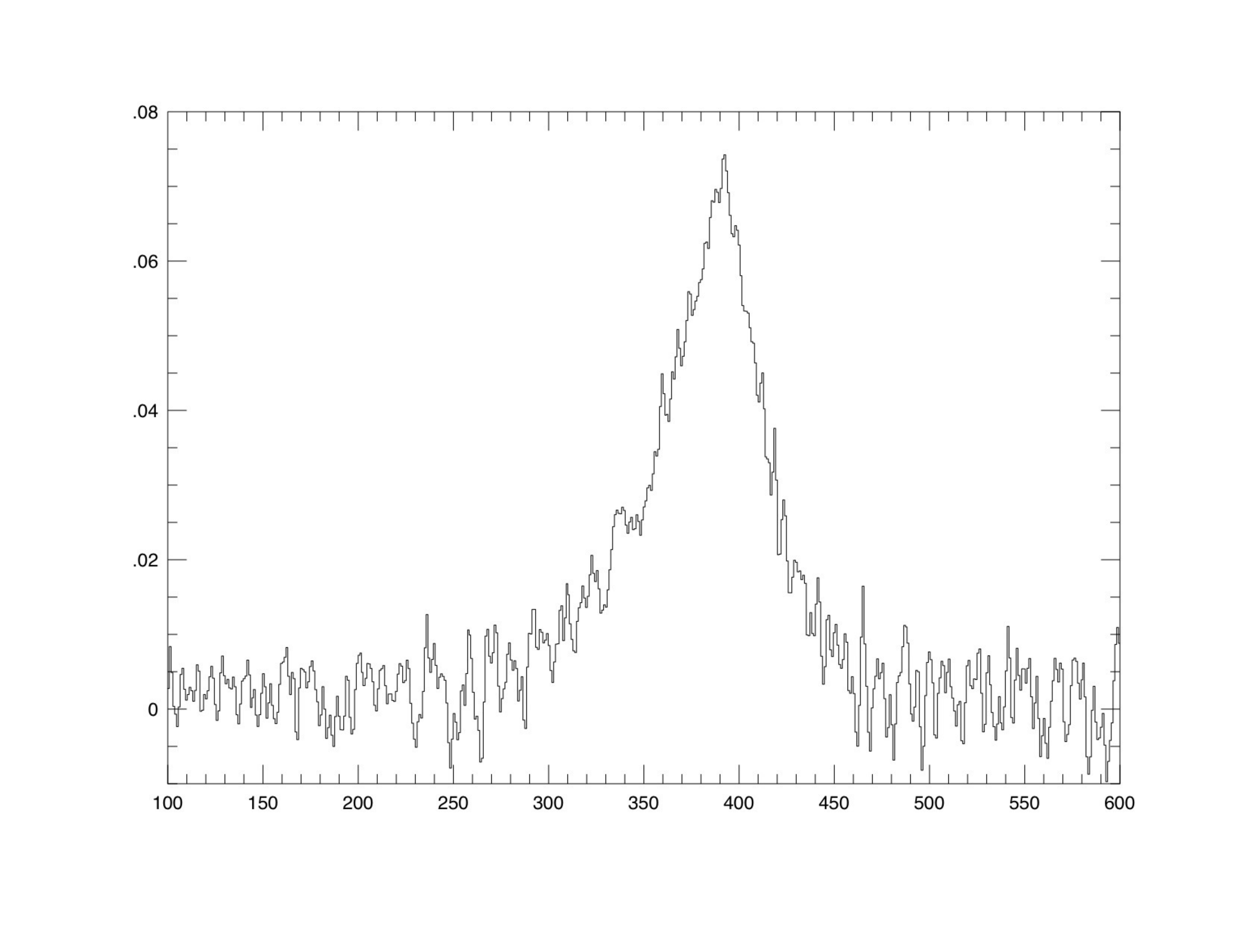}{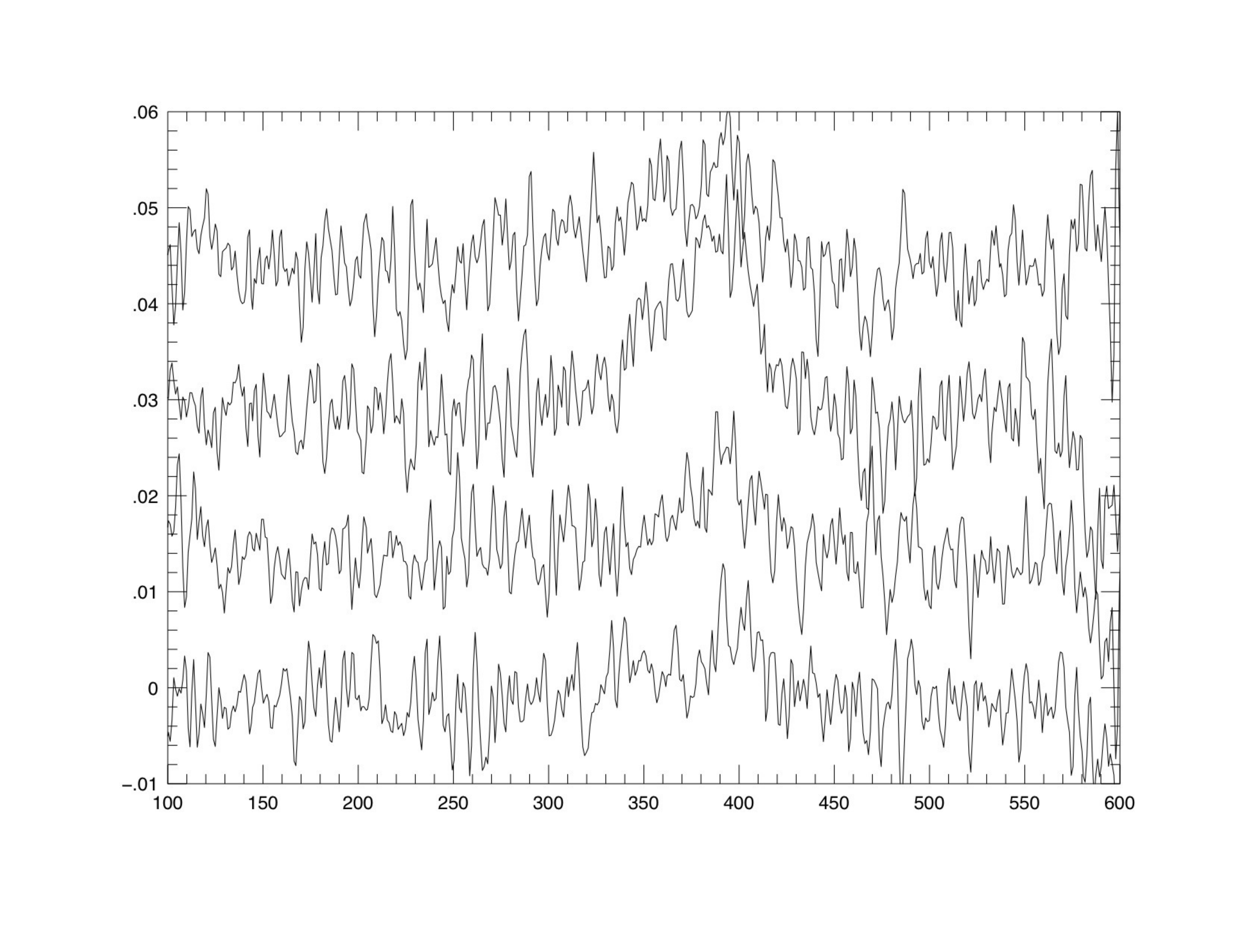}
\caption{Top:  The spectrum in the peak  position averaged over $3\arcsec$ along the slit. Intensity units are ${\rm erg(s~cm^{-1}~cm^2~sr)^{-1}}$; x axis is heliocentric velocity. Bottom:  Spectra in the weaker positions averaged over $3\arcsec$  along the slit. The spectra are, from top to bottom,  $1\arcsec$ east, $1\arcsec$, $2\arcsec$, and $3\arcsec$ west of the peak.  Spectra are offset vertically by steps of  0.015. Units as in previous figure.}
\end{center}
\end{figure}

\begin{figure}
\begin{center}
\includegraphics[scale=0.3]{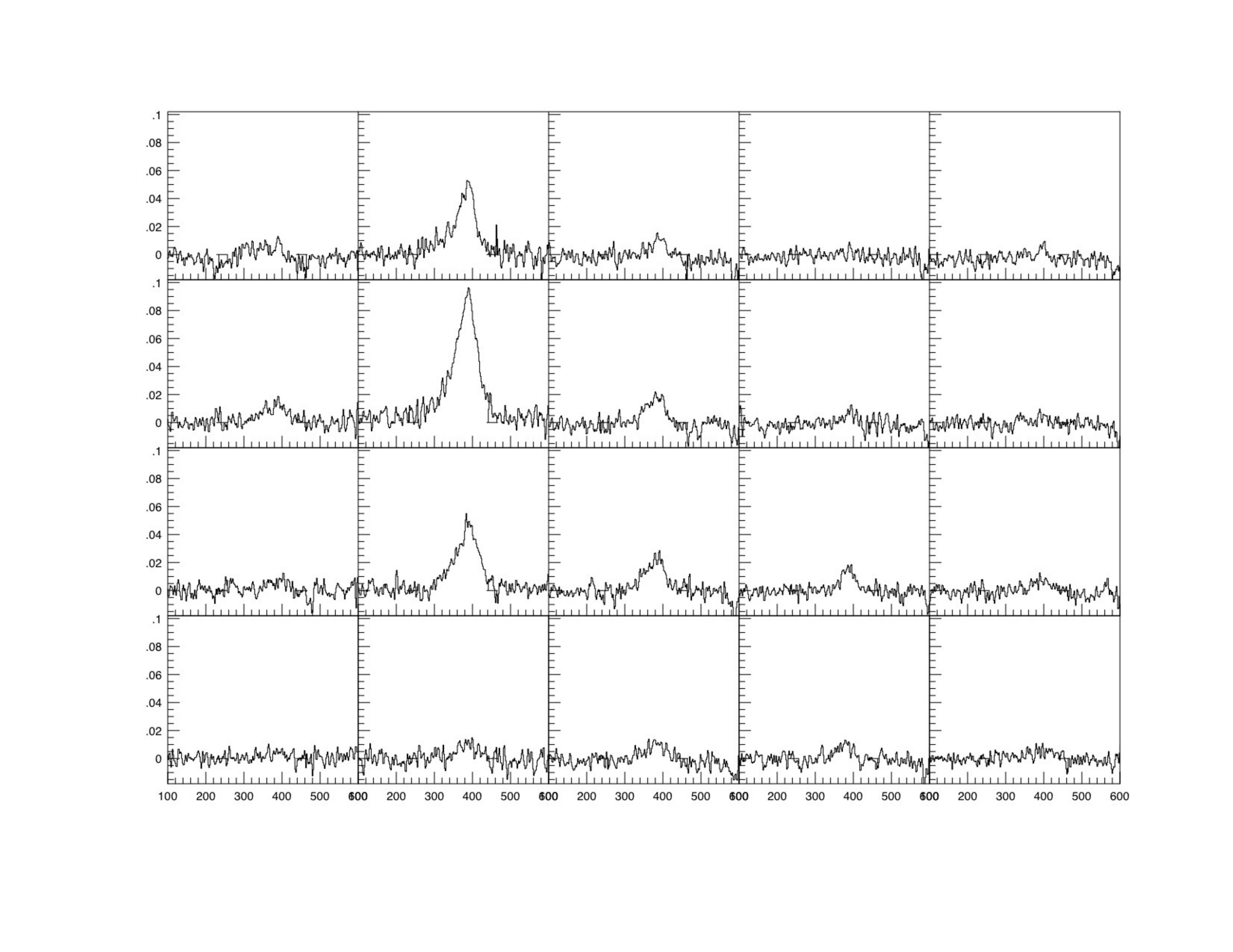}
\caption{The spectrum of [S\,IV] at each point on a $1\arcsec \times 1.1\arcsec$ grid. North is up and East left. The data were binned by 3 pixels along the slit and 2 pixels in velocity.  Units as in Figure 1. }
\end{center}
\end{figure} 

\begin{figure}
\begin{center}
\includegraphics[scale=0.4]{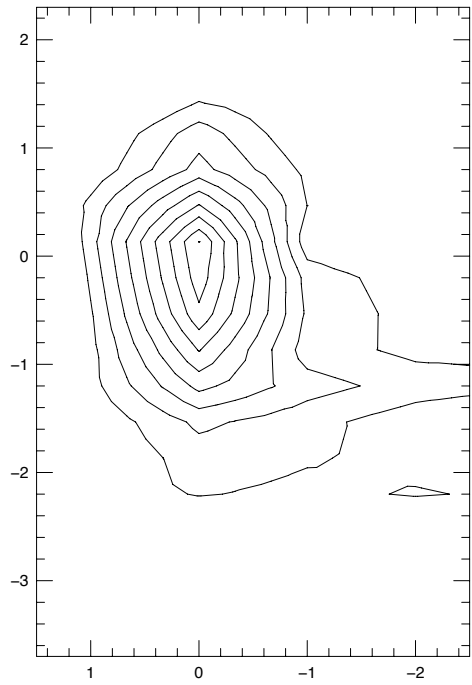}
\caption{[S\,IV] line intensity integrated over +275 -- +465\kms , showing the spatial distribution of the line. North is up, East is left.    Contours are integer multiples of ${\rm1.6\times10^{-2}erg(s~cm^{2}~sr)^{-1}}$. }
\end{center}
\end{figure}

\begin{figure}
\begin{center}
\includegraphics[scale=1.0]{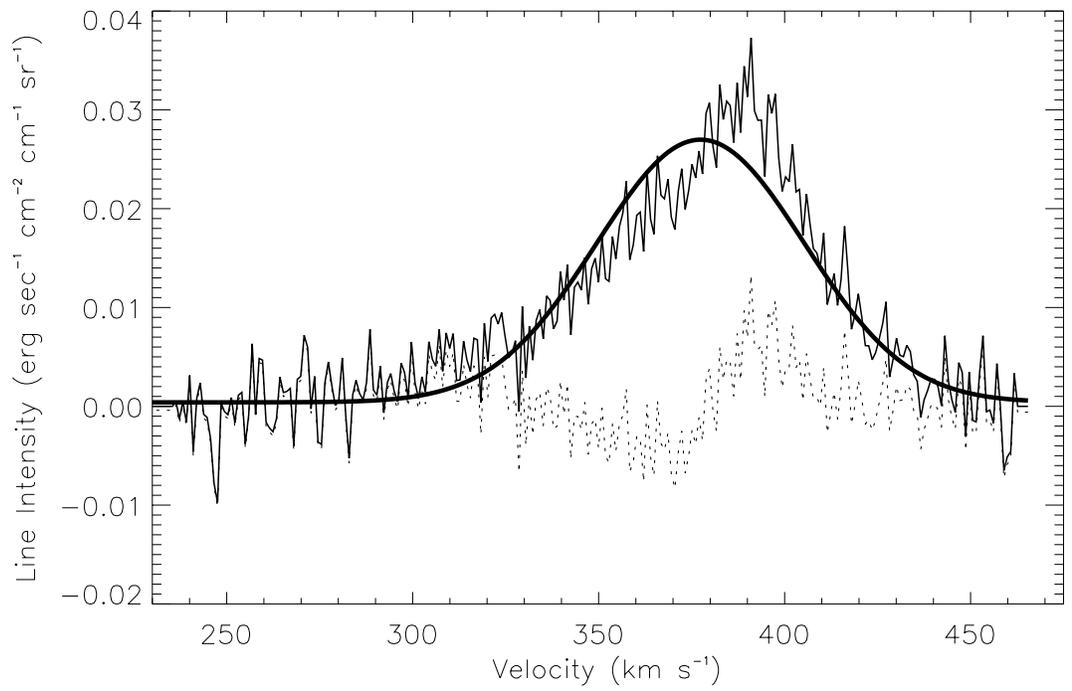}
\caption{The [S\,IV] line observed in the strongest position, the best fitting single Gaussian, and the residuals.   }
\end{center}
\end{figure}

\begin{figure}
\begin{center}
\includegraphics[scale=1.0]{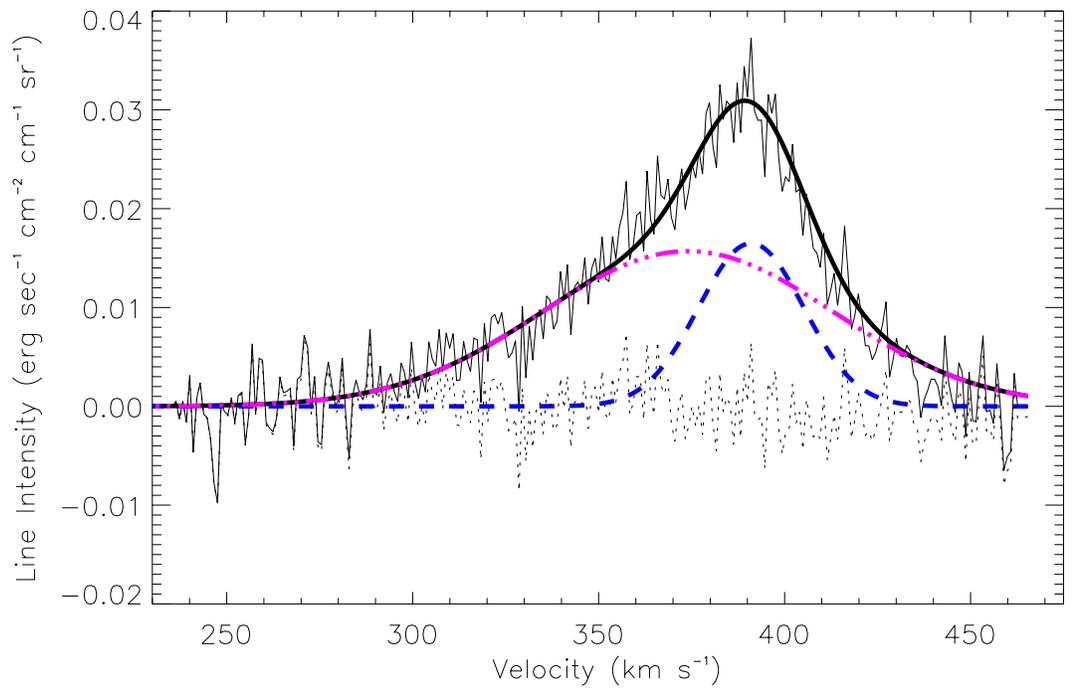}
\caption{The [S\,IV] line observed in the strongest position, the best fitting sum of two Gaussians, and the residuals.}
\end{center}
\end{figure}

\begin{figure}
\begin{center}
\includegraphics{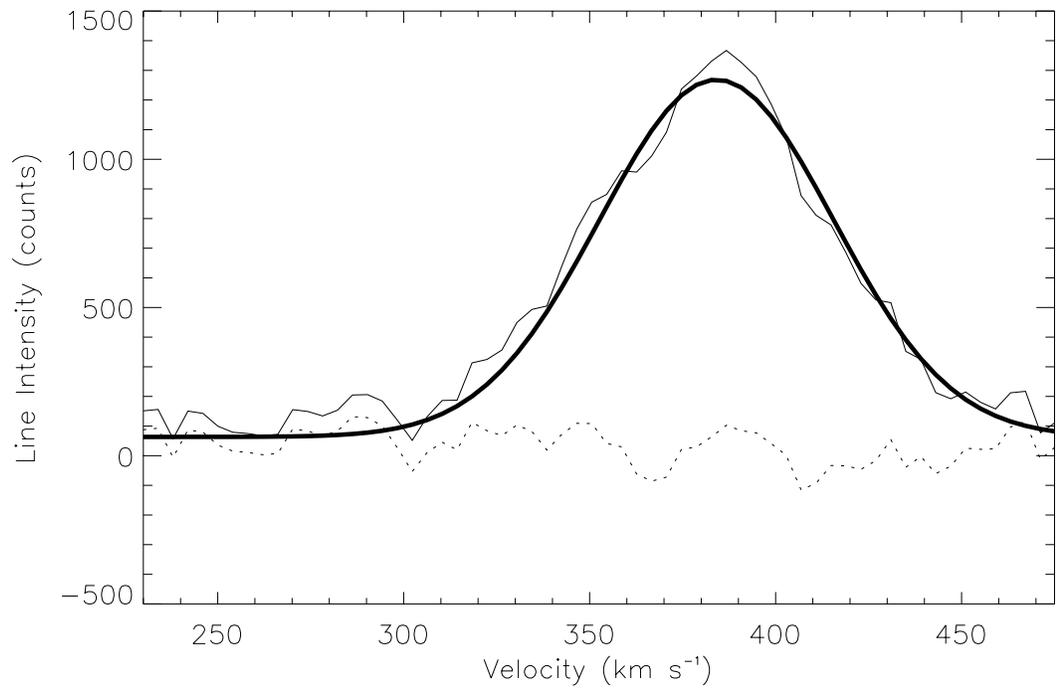}
\caption{The \bra\ line observed with 12\kms\ resolution. The data is from  \citet{TB03} but taken over a 1.2\arcsec\ slit. Also shown are the best single Gaussian fit to the line and the residuals of the fit.  Y-axis units are counts and velocity is LSR.}
\end{center}
\end{figure}

\begin{figure}
\begin{center}
\includegraphics{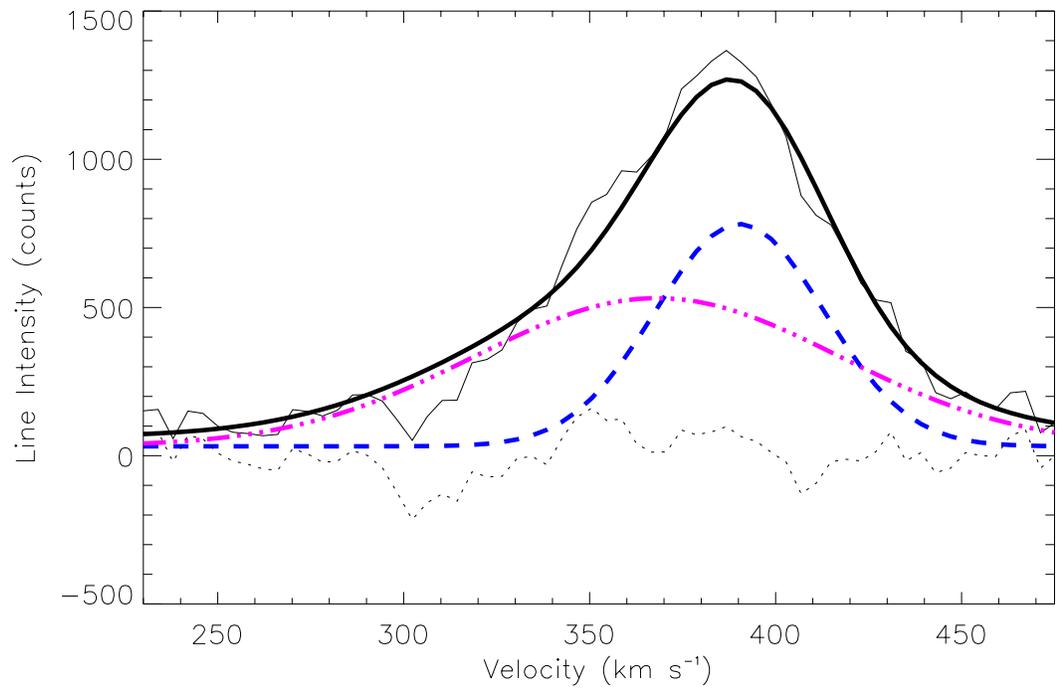}
\caption{The \bra\ line of \citet{TB03}, the best two Gaussian fit and the residuals to the fit.  For this line, in contrast to the [S\,IV], the two Gaussian fit is not clearly better than one Gaussian.}
\end{center}
\end{figure}

\begin{figure}
\begin{center}
\includegraphics[scale=0.7]{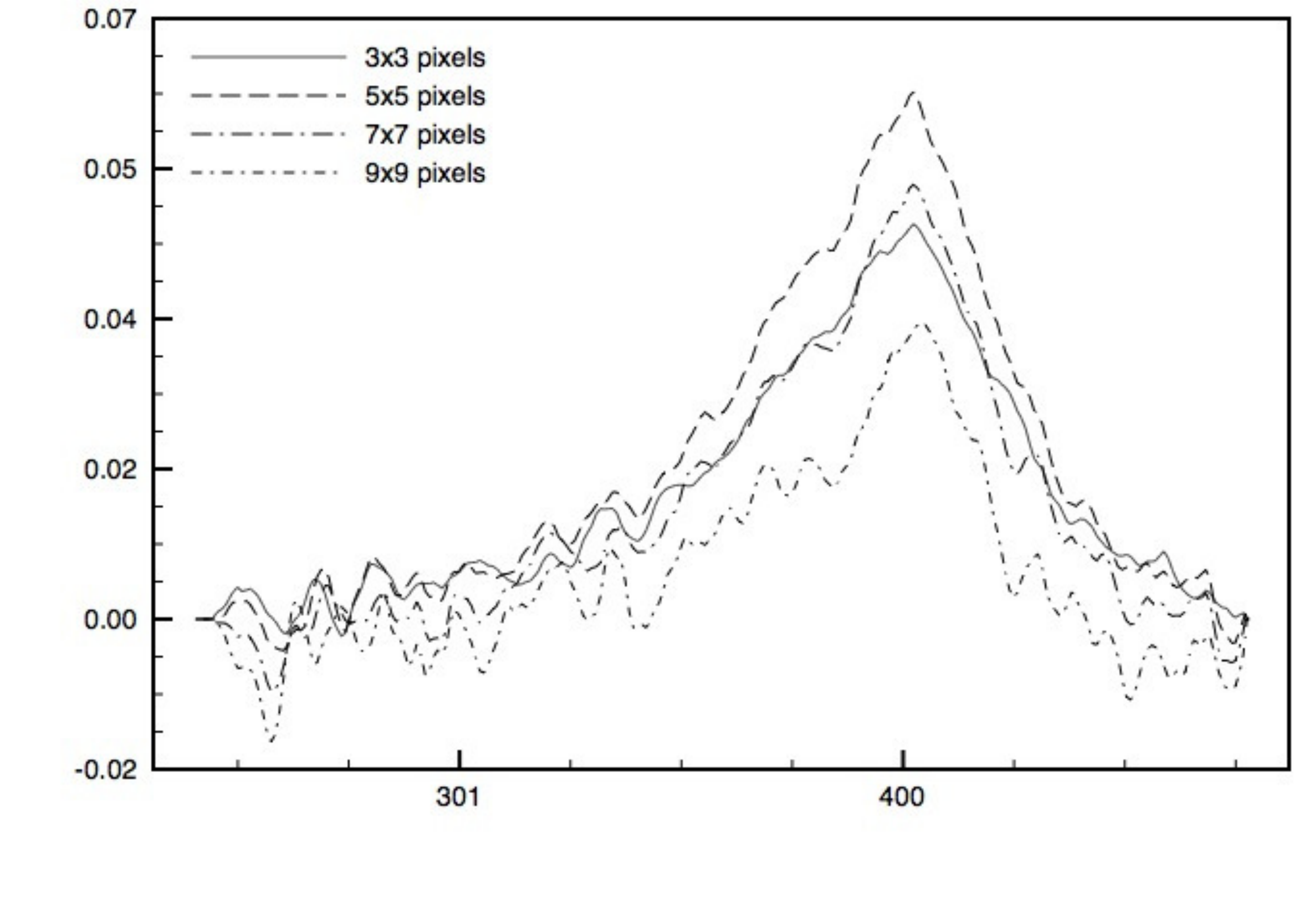}
\caption{The [S\,IV] line in boxes or annuli around the peak. The boxes are $3 \times 3$, $5 \times 5$, $7 \times 7$ and $9 \times 9$ pixels.  The smaller boxes are subtracted from the larger ones, i.e. the $5 \times 5$ pixel spectrum does not contain contributions from the inner $3 \times 3 $ pixel box.  Units are as in Figure 1.}
\end{center}
\end{figure}


\begin{thebibliography}{}
\bibitem[Alonso-Herrero et al.(2004)]{AAH04}
Alonso-Herrero, A., Takagi,T., Baker, A., Rieke, G., Rieke, M., Imanishi, M, \& Scoville, N. \ 2004, \apj, 612, 222 
\bibitem[Beck (2008)]{B08}
Beck, S.C. 2008, \aap. 489, 567 
\bibitem[Beck et al.(1996)]{BL96}
Beck, S.C., Turner, J.L., Ho, P.T.P., Lacy, J., \& Kelly, D.M. \ 1996, \apj, 467, 610 
\bibitem[Calzetti et al.(1997)]{1997AJ....114.1834C} 
Calzetti, D., Meurer, G.~R., Bohlin, R.~C., Garnett, D.~R., Kinney, A.~L., Leitherer, C., 
\& Storchi-Bergmann, T.\ 1997, \aj, 114, 1834 
\bibitem[Calzetti et al.(1999)]{1999AJ....118..797C} 
Calzetti, D., Conselice, C.~J., Gallagher, J.~S., III, \& Kinney, A.~L.\ 1999, \aj, 118, 797 
\bibitem[Crowther et al (1999)]{CB99}
Crowther P.A., Beck S.C., Willis A.J., Conti P.S., Morris P.W.,
Sutherland R.S. \ 1999, \mnras, 304, 654 
\bibitem[Crowther et al (2010)]{CR10}
Crowther, P., Schnurr, O., Hirschi, R., Yusof, N., Parker, R., Goodwin, S.,  \& Kassim, H. \ 2010, \mnras, 408, 731
\bibitem[Donahue \& Shull (1991)]{DS91}
Donahue, M. \& Shull, J. M. \ 1991, \apj, 383, 511
\bibitem[Gilbert \& Graham (2007)]{GG07}
Gilbert, A. M., \& Graham, J. \ 2007, \apj, 668, 168 
\bibitem[Graham(1981)]{1981PASP...93..552G} Graham, J.~A.\ 1981, \pasp, 93,
552
\bibitem[Harris et al.(2004)]{2004ApJ...603..503H} 
Harris, J., Calzetti, D., Gallagher, J.~S., III, Smith, D.~A., \& Conselice, C.~J.\ 2004, \apj, 603, 503
\bibitem[Henry et al.(2007)]{2007AJ....133..757H} Henry, A.~L., Turner, 
J.~L., Beck, S.~C., Crosthwaite, L.~P., 
\& Meier, D.~S.\ 2007, \aj, 133, 757 
\bibitem[Jaffe et al. (2003)]{DJ03}
Jaffe, D., Zhu, Q., Lacy, J.H., \& Richter, M. \ 2003. \apj, 596, 1053
\bibitem[Kobulnicky \& Skillman(2008)]{2008AJ....135..527K} 
Kobulnicky, H.~A., \& Skillman, E.~D.\ 2008, \aj, 135, 527 
\bibitem[Kobulnicky et al.(1997)]{1997ApJ...477..679K} Kobulnicky, H.~A., 
Skillman, E.~D., Roy, J.-R., Walsh, J.~R., 
\& Rosa, M.~R.\ 1997, \apj, 477, 679 
\bibitem[Kroupa \& Boily(2002)]{2002MNRAS.336.1188K} 
Kroupa, P., \& Boily, C.~M.\ 2002, \mnras, 336, 1188 
\bibitem[Kurtz (2005)]{SK05}
Kurtz, S. , Proceedings IAU Symposium 227: Massive Star Birth, eds. Cesaroni, Felli, Churchwell \& Walmsley (2005)
\bibitem[Lacy et al.(2002)]{LA02}
Lacy, J., Richter, M., Greathouse, T., \& Zhu, Q-F. \ 2002, \pasp, 114, 153 
\bibitem[L{\'o}pez-S{\'a}nchez et al.(2007)]{LS07}
L{\'o}pez-S{\'a}nchez, {\'A}.~R., Esteban, C., Garc{\'{\i}}a-Rojas, J.,
Peimbert, M., \& Rodr{\'{\i}}guez, M.\ 2007, \apj, 656, 168
\bibitem[Mart{\'{\i}}n-Hern{\'a}ndez et 
al.(2005)]{2005A&A...429..449M} 
Mart{\'{\i}}n-Hern{\'a}ndez, N.~L., Schaerer, D., \& Sauvage, M.\ 2005, \aap, 429, 449 
\bibitem[Meatheringham \& Dopita (1991)]{MD91}
Meatheringham, S. \& Dopital, M. \ 1991 \apjs, 75, 407 
\bibitem[Meier et al.(2002)]{MT02} Meier, D.~S., Turner, 
J.~L., \& Beck, S.~C.\ 2002, \aj, 124, 877 
\bibitem[Meurer et al (1995)]{MH95}
Meurer, G.R., Heckman, T.M., Leitherer, C., Kinney, A., Robert, C. \& Garnett, D.R. \ 1995, \aj, 110, 2665 
\bibitem[Mohan et al.(2001)]{2001ApJ...557..659M} Mohan, N.~R., 
Anantharamaiah, K.~R., \& Goss, W.~M.\ 2001, \apj, 557, 659 
\bibitem[Monreal-Ibero et al.(2010)]{MI10} 
Monreal-Ibero, A., V{\'{\i}}lchez, J.~M., Walsh, J.~R., \& Mu{\~n}oz-Tu{\~n}{\'o}n, C.\ 2010, \aap, 517, A27 
\bibitem[Persson et al (1984)]{PG84}
Persson, E., Geballe, T.R., McGregor, P.J., Edwards, S., \& Lonsdale, C.J. \ 1984, \apj, 286, 289 
\bibitem[Schaerer et al.(1997)]{1997ApJ...481L..75S} Schaerer, D., Contini, 
T., Kunth, D., \& Meynet, G.\ 1997, \apjl, 481, L75 
\bibitem[Rodriguez-Rico et al (2007)]{RG07}
Rodriguez-Rico, C. A.; Goss, W. M.; Turner, J. L.; Gomez, Y. \ 2007, \apj, 670, 295 
\bibitem[Tobin et al (2009)]{TO09}
Tobin, J., Hartmann, L., Furesz, G., Mateo, M., \& Megeath, S. \ 2009, \apj, 697, 1103 
\bibitem[Tremonti et al.(2001)]{2001ApJ...555..322T} Tremonti, C.~A., 
Calzetti, D., Leitherer, C., \& Heckman, T.~M.\ 2001, \apj, 555, 322 
\bibitem[Turner et al.(1997)]{TBH97} Turner, J.~L., Beck,
S.~C., \& Hurt, R.~L.\ 1997, \apjl, 474, L11
\bibitem[Turner et al.(1998)]{1998AJ....116.1212T} Turner, J.~L., Ho, 
P.~T.~P., \& Beck, S.~C.\ 1998, \aj, 116, 1212 
\bibitem[Turner et al.(2000)]{2000ApJ...532L.109T} Turner, J.~L., Beck, 
S.~C., \& Ho, P.~T.~P.\ 2000, \apjl, 532, L109 
\bibitem[Turner et al.(2003)]{TB03}
Turner, J. L.; Beck, S. C.; Crosthwaite, L. P.; Larkin, J. E.; McLean, I. S.; Meier, D. S. \ 2003 \nat, 423, 621 
\bibitem[Turner \& Beck (2004)]{TB04}
Turner, J.\& Beck, S.C. \ 2004, \apjlett,  602, ,85L 
\bibitem[Underhill (1969)]{UN69}
Underhill, A.  \ 1969, \apss 3, 109 
\bibitem[Vanzi et al.(2006)]{2006A&A...459..769V} 
Vanzi, L., Scatarzi, A., Maiolino, R., \& Sterzik, M.\ 2006, \aap, 459, 769 
\bibitem[Zastrow et al.(2011)]{2011ApJ...741L..17Z} Zastrow, J., Oey,
M.~S., Veilleux, S., McDonald, M., \& Martin, C.~L.\ 2011, \apjl, 741, L17
\bibitem[Zhu et al. (2008)]{ZL08}
Zhu, Q.-F., Lacy, J.H., Jaffe, D.T., Richter, M.J., \& Greathouse, T.K., \ 2008 \apjs, 177, 584 
\bibitem[Zhu (2006)]{ZH}
Zhu, Q.-F. \ 2006, PhD Thesis UT Austin, UMI 3246104 

\end{thebibliography}
\end{document}